\documentclass[sn-nature]{sn-jnl}

\usepackage{graphicx}%
\usepackage{multirow}%
\usepackage{amsmath,amssymb,amsfonts}%
\usepackage{amsthm}%
\usepackage{mathrsfs}%
\usepackage[title]{appendix}%
\usepackage{xcolor}%
\usepackage{textcomp}%
\usepackage{manyfoot}%
\usepackage{booktabs}%
\usepackage{algorithm}%
\usepackage{algorithmicx}%
\usepackage{algpseudocode}%
\usepackage{listings}%

\theoremstyle{thmstyleone}%
\theoremstyle{thmstyletwo}%

\theoremstyle{thmstylethree}%

\begin{document}

\title[Article Title]{How Dark is Dark Energy?}


\author[1,2,3]{\fnm{Mauro} \sur{Carfora}}\email{mauro.carfora@unipv.it}

\author[1,2,3]{\fnm{Francesca} \sur{Familiari}}\email{francesca.familiari01@universitadipavia.it}


\affil[1]{\orgdiv{Department of Physics}, \orgname{University of Pavia}, \orgaddress{\street{Via Bassi 6}, \city{Pavia}, \postcode{27100}, \country{Italy}}}

\affil[2]{\orgname{National Institute for Nuclear Physics (INFN)-Pavia Section}}

\affil[3]{\orgname{National Group for Mathematical Physics (GNFM)-Pavia Unit}}

\abstract{The nature of dark energy is one of the fundamental problems in cosmology. Introduced to explain the apparent acceleration of the Universe's expansion, its origin remains to be determined. In this paper, we illustrate a result that may significantly impact understanding the relationship between dark energy and structure formation in the late-epoch Universe. Our analysis exploits a scale-dependent energy functional, initially developed for image visualization, to compare the physical and geometrical data that distinct cosmological observers register on their celestial spheres. In the presence of late-epoch gravitational structures, this functional provides a non-perturbative technique that allows the standard Friedmann-Lema\^itre-Robertson-Walker (FLRW) observer to evaluate a measurable, scale-dependent difference between the idealized FLRW past light cone and the physical light cone. From the point of view of the FLRW observer, this 
 difference manifests itself as a redshift-dependent correction $\Lambda^{(corr)}(z)$ to the FLRW cosmological constant $\Lambda^{(FLRW)}$. At the scale where cosmological expansion couples with the local virialized dynamics of gravitational structures, we get
$\Lambda^{(corr)}(z)\sim 10^{-52}\,m^{-2}$, indicating that the late-epoch structures induce an effective cosmological constant that is of the same order of magnitude as the assumed value of the FLRW cosmological 
constant, a result that may lead to an interpretative shift in the very role of dark energy. }




\maketitle

\section{Introduction}\label{sec1}
Dark energy, in the form of a cosmological constant $\Lambda$, cold dark matter (CDM), and inflation are the key players in the $\Lambda\mathrm{CDM}$ model, the standard paradigm of modern cosmology. A rich repertoire of hypotheses, modeling assumptions, and observations\cite{Planck} indicate that this dark backstage is responsible for most of the large-scale physics in our Universe, from the early stage of structure formation to the observed late-epoch acceleration of the cosmological expansion. We are familiar with the complex astrophysical landscape that dominates our neighborhood,  yet the Universe appears statistically\cite{SEH} isotropic and homogeneous when observed on large scales. This cosmic view embraces space and time, and the Universe was indeed extremely uniform, up to tiny density fluctuations when the cosmic microwave background (CMB) radiation formed. Surveys of the CMB indicate that these primordial density fluctuations have a Gaussian, almost scale-invariant power spectrum originating from inflation and suggest that the inhomogeneities around us are the gravitational evolution of these inflationary seeds. We are dealing with a top-down scenario that is difficult to verify in a model-independent way\cite{Maartens}, but which is instrumental in describing cosmological dynamics in terms of a Friedmann-Lema\^itre-Robertson-Walker (FLRW) spacetime (with flat space sections) and its perturbations. The FLRW evolution of the $\Lambda\mathrm{CDM}$ model depends quite rigidly on a minimal set of observational data, the most relevant of which are the present value of the Hubble expansion parameter,  the matter density (comprising baryonic and dark matter),\; the radiation density, the primordial fluctuation amplitude related to inflation, and the scalar spectral index characterizing how fluctuations change with the scale. \; From the Friedmann equations, derived from Einstein's theory, we eventually get also the presence of a dark energy contribution in the vague terms of a cosmological constant.\;  Observations show that, at the present epoch, the estimated value of this dark energy contribution is dominant over the other components, a dominance that we see at work in the actual accelerated phase of the cosmological expansion. In a nutshell, this narrative is the $\Lambda\mathrm{CDM}$ concordance model of cosmology. Simple, somewhat vague, surprisingly predictive\cite{Planck}. Yet, the simplicity of this description is deceptive since we live in a perturbed universe that bears, in a rather unexpected and subtle way, the imprint of the fluctuations induced by gravitational instability. The successful role of  FLRW spacetime geometry and its perturbations
in the early-epoch $\Lambda\mathrm{CDM}$ scenario, is related to the observed transition to homogeneity\cite{Scrimgeour, Dias} that occur for comoving radii of the order $100\mathrm{h}^{-1}\; \mathrm{Mpc}$, where $\mathrm{h}$ is the dimensionless parameter describing the relative uncertainty of the present value, $H_0$, of the Hubble parameter.
It is important to stress that the characterization of this homogeneity scale depends on the statistical measure used, typically two-point correlation function statistics that exploit an FLRW background as a reference. Model-independent statistics containing all orders of the correlation function push the transition to homogeneity at larger scales. Here, we do not belabor on this delicate issue, and for simplicity, we use the standard reference value $100\mathrm{h}^{-1}\; \mathrm{Mpc}$ adopted in mainstream cosmology. 
In the late-epoch Universe, the role of FLRW geometry in the $\Lambda\mathrm{CDM}$ model is more delicate to handle. Gravitational clustering generates fluctuations that grow larger and larger and eventually become non-perturbative. Observations show that, in the pre-homogeneity region surrounding us, gravitational clustering gives rise to a complex network of structures, characterized by a foam-like web of voids and galaxy filaments often extending well into the  $100\; \mathrm{Mpc}$ range\cite{Wiltshire}. At these scales,  the Einstein evolution of the fiducial  FLRW geometry uncouples from the dynamics of the matter sources and survives more as a practical computational assumption, often only assisted by  Newtonian and recently general-relativistic, (model-dependent) numerical simulations, rather than as a  \emph{bona fide} perturbative background gravitationally determined by the actual matter distribution as  Einstein's theory requires. We have little perturbative control over spacetime geometry in such pre-homogeneity regions. In particular,  the transition from their large-scale FLRW point of view to the actual inhomogeneous and anisotropic spacetime geometry emergent at these local scales is poorly understood in a model-independent way, and the idea that around $100\mathrm{h}^{-1}\; \mathrm{Mpc}$ we have a gradual and smooth transition between these two regimes is somewhat illusive. The existence of this issue was presciently put forward long ago by G. F. R. Ellis \cite{EllisPadova} and explored in depth by T. Buchert\cite{DarkBuchert, BuchertSyksy} (there is a vast literature on the subject, see\cite{Ellis2} for a review). Ellis suggested that non-perturbative inhomogeneities can generate backreaction effects due to the nonlinear nature of Einstein equations. These effects can contribute to the overall FLRW dynamics by dressing the cosmological parameters and even motivate a perspective shift on the nature of dark energy\cite{DarkBuchert, Carfora0, Carfora1}. The lively debate raised by backreaction\cite{BuchertSyksy, Buchert, Ellis2, Wald1, Wald2, Wald3}, and an in-depth analysis of nonlinear FLRW perturbative effects\cite{Durrer, Fanizza, Heinesen, Obinna4} showed that backreaction could affect cosmological parameters up to the percent level, and it has a role in high-precision cosmology. But what about the impact of inhomogeneities\cite{Durrer} and backreaction on dark energy? Evidence in this sense is not conclusive. Perturbative results and averaging techniques leave the solution of the problem in an unclear state. There are rigidities to address: the problem is not perturbative but dynamical. Observations show that the expansion of the Universe appears to have started accelerating at the same epoch when complex nonlinear structures emerged, a dynamical evolution that is technically difficult to control with the available backreaction and perturbative methods.  
\\
The primary motivation for this paper is to illustrate a new non-perturbative approach and a result that may significantly impact understanding the role of inhomogeneities in the late-epoch Universe. We work within mainstream cosmology,  introducing a scale-dependent comparison functional between light cones\cite{CarFam} to evaluate the difference between the idealized FLRW past light cone and the physical light cone associated with the actual cosmological data. This comparison functional is naturally defined on the celestial sphere of the FLRW observer and can be directly related to the fluctuations between the physical and the FLRW area distances. As such, it is an observable quantity that, at the given scale, readily informs the FLRW observer of the global difference between her light cone description and the one provided by the physical light cone. 
As we approach the \emph{cosmological coupling transition} in the pre-homogeneity region, where structures, gravitationally virialized, start coupling to the cosmological expansion, this global difference manifests itself as a scale-dependent correction to the FLRW cosmological constant. The correction is so significant that it may lead to an interpretative shift in dark energy.

\section{A tale of two observers} 
\begin{figure}[h]
\centering
\includegraphics[scale=0.3]{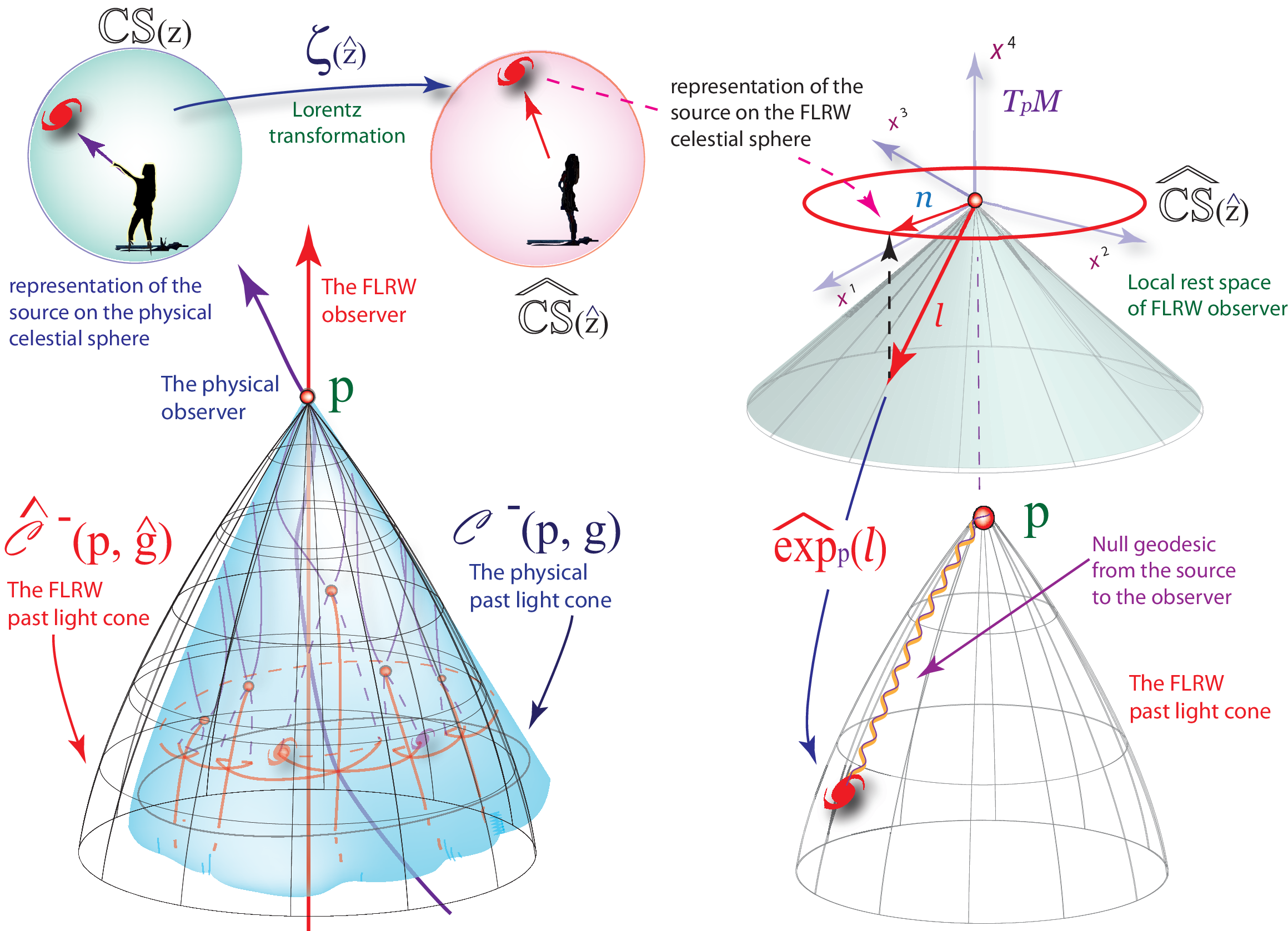}
\caption{A pictorial rendering of the relation between the physical observer and her FLRW avatar. They share the observational event $p$ but generally have distinct 4-velocities. On the left part of the drawing, there is a representation of the light cones associated with the two observers and their respective celestial spheres. The right side of the picture provides a characterization of the FLRW celestial sphere (represented as a circumference for dimensional reason) in the local rest space $T_pM$ of the observational event (the tangent space to the FLRW spacetime $M$ at $p$).  The vector $\textbf{n}$ defines the spatial direction pointing at the source as seen by the FLRW observer. The null vector $\textbf{l}$ is the associated past-directed null direction. The corresponding past-directed null geodesic reaching the source on the FLRW past light cone is described using the exponential map $\widehat{\exp}_p$, the map that to a (null) vector $\textbf{l}\in T_pM$ associates the corresponding spacetime null geodesic. A similar characterization holds for the physical celestial sphere.}
\label{<fig2>}   
\end{figure}
According to the best-fit point of view advocated long ago\cite{EllisStoeger} by G. F. R. Ellis and W. R. Stoeger, the actual Universe should be modeled by a physical spacetime manifold $M$ with a Lorentzian metric $g$ that is statistically isotropic and homogeneous, \emph{viz.}\, a perturbed FLRW,  only on scales greater than the scale $100\mathrm{h}^{-1}\; \mathrm{Mpc}$ that marks the transition to homogeneity\cite{Gott, Hogg, Scrimgeour}. This assumption implies that on large scales $(M, g)$ can be (perturbatively) described in terms of a background FLRW spacetime $(M, \widehat{g})$. On smaller scales and late-epoch, the description of $(M, g)$ is more complex  since, according to the Einstein equations, the spacetime geometry must comply with the network of structures that dominate the cosmological landscape for  
$\lesssim\,100\mathrm{h}^{-1}\; \mathrm{Mpc}$. In this pre-homogeneity region, the typical observer is associated with an event $p$, the \emph{here and now} corresponding to which she gathers cosmological data. In practice, the \emph{here} can be identified with the barycenter of our small cluster of galaxies (the Local Group), and the \emph{now} is characterized by the actual temperature $T_{CMB}\,=\,2.725$ of the cosmic microwave background (CMB) as measured in the frame centered on us but stationary with respect to the CMB.   The interpretation of data in the pre-homogeneity region surrounding us is complex since the FLRW model should be used only over the largest scales. Yet, FLRW observers and the associated FLRW spacetime $(M, \widehat{g})$ have a natural role also in the presence of inhomogeneities as advocated by E. Kolb, V. Marra, and S. Matarrese\cite{KolbMarraMatarrese}.
Thus, together with the physical observer, it is natural to consider her Friedmannian avatar, a reference FLRW observer located at $p$ and possibly moving, with respect to the physical observer, with a relative velocity $v(p)$. The two observers may compare the geometry that astrophysical data induce on the celestial spheres associated with the physical past light cone, $\mathcal{C}^-(p, g)$, and the FLRW past light cone, $\mathcal{C}^-(p, \widehat{g})$,   see Fig.\ref{<fig2>}. The physical celestial sphere,\; $\mathbb{CS}_z$, \, is defined by the set of (null) directions that point, in the instantaneous rest space of the observer, to the astrophysical sources at the given redshift $z$. Similarly  $\widehat{\mathbb{CS}}_{\widehat{z}}$ defines the celestial sphere, at redshift $\widehat{z}$, of the reference FLRW observer. Since the observers may  have different 4-velocities, these two celestial spheres are identified modulo a Lorentz transformation
\begin{equation}
\zeta_{(_{\widehat{z}})}\,:\,\widehat{\mathbb{C\,S}}_z\,\longrightarrow\,\mathbb{C\,S}_{\widehat{z}}
\end{equation}
determined by the images, on $\mathbb{C\,S}_{\widehat{z}}$, of three distinct astrophysical sources of choice (\emph{e.g.} Cepheid variable stars) on the FLRW celestial sphere $\widehat{\mathbb{C\,S}}_{\widehat{z}}$. By optimizing the choice of these three astrophysical sources on 
$\widehat{\mathbb{C\,S}}_{\widehat{z}}$, the FLRW observer can relate the redshifts ${\widehat{z}}$ and $z$ (optimization is necessary since the sources are typically affected by peculiar motions with respect to the reference FLRW Hubble flow). For astronomical descriptive purposes, celestial spheres are usually described as round two-spheres, celestial cartography globes that depict the location of sources in the observer's sky at the given redshift. But there is a more accurate rendering of this celestial cartography, a rendering analogous to the description of the Earth's surface in terms of a globe with raised relief to show mountains and landforms.

\section{The Method:\,Comparing celestial spheres} 

\begin{figure}[h]
\centering
\includegraphics[scale=0.3]{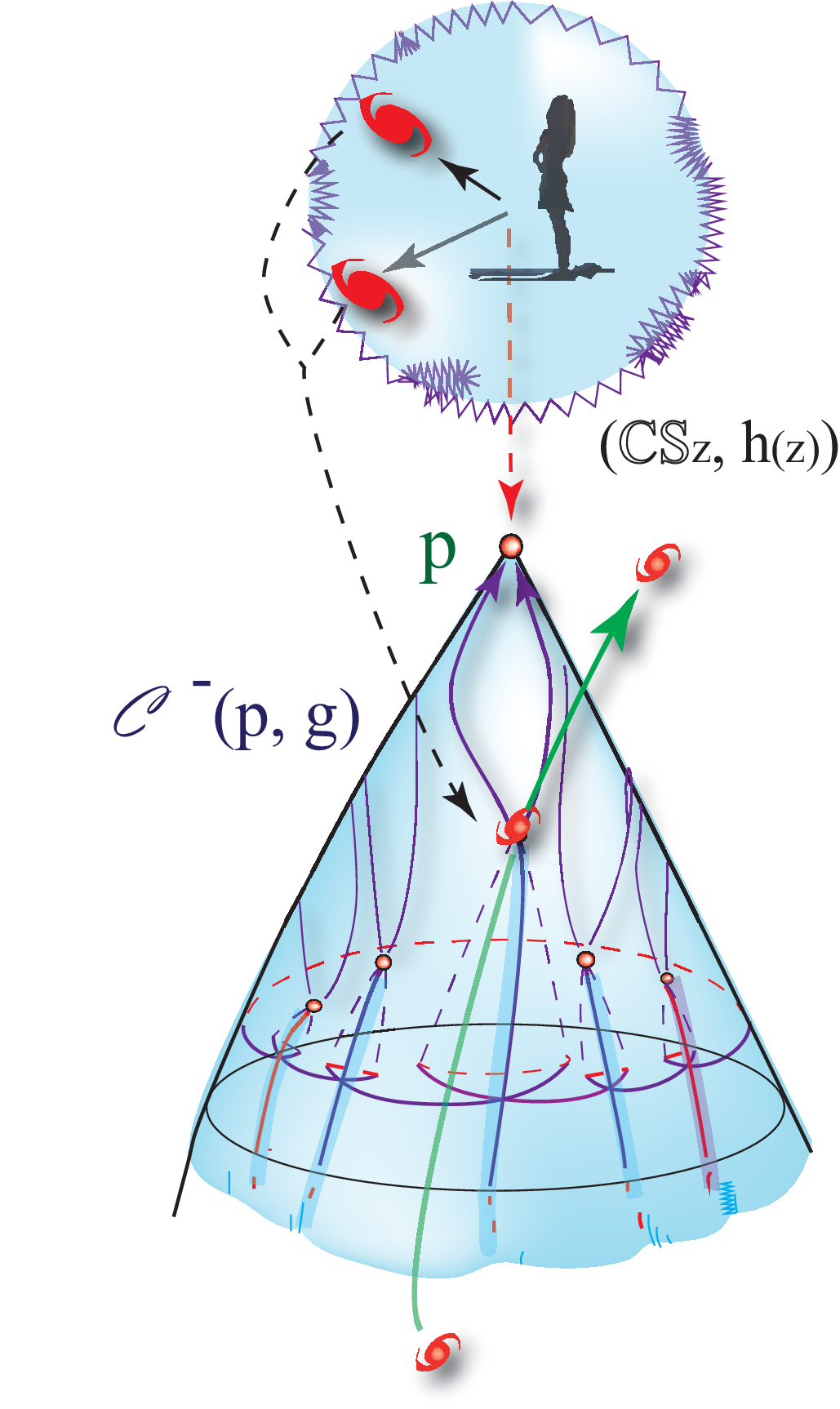}
\caption{In the pre-homogeneity region, null geodesics develop conjugate and cut locus points with the ensuing formations of past light cone caustics. On the physical observer's celestial sphere $\mathbb{CS}_z$, this behavior gives rise to the various distortions familiar in gravitational lensing. Here, for instance, we have a pictorial rendering of the apparent double image of a single galaxy. Globally, these distortions take form in the metric geometry $h_{(z)}$ that the null geodesic flow induces (via the exponential map $\exp_p$) on $\mathbb{CS}_z$. The figure describes this geometry as the jagged celestial sphere; we can profitably handle its properties with the Lipschitzian  methods described in \cite{CarFam, CarFam2} .}
\label{<fig3>}   
\end{figure}

\begin{figure}[h]
\centering
\includegraphics[scale=0.3]{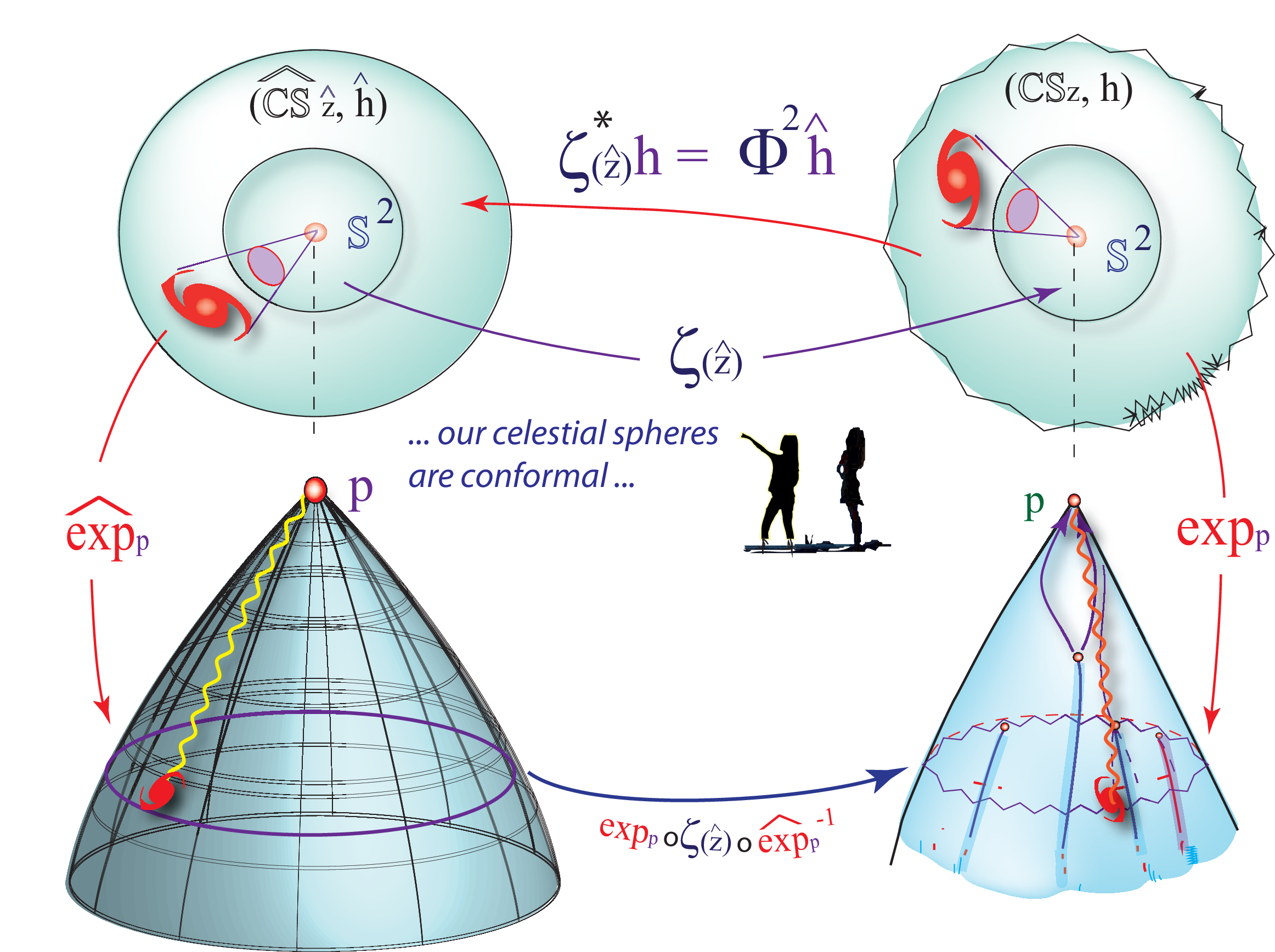}
\caption{The construction of the conformal mapping between the physical and the FLRW celestial spheres directly results from characterizing the celestial spheres and the Lorentz transformation compensating for the observer's relative velocity. The picture shows the maps involved in this construction; their joint action (see the bottom of the drawing) acts between (constant redshift) sections of the light cones, an action that, under natural physical assumptions, characterizes the conformal transformation of the associated celestial spheres. }
\label{<fig5>}   
\end{figure}

Data from astrophysical sources at a given redshift,
reach the physical observer and are portrayed on her celestial sphere $\mathbb{CS}_z$, by traveling along null geodesics. Physical null geodesics are not smooth since they may develop conjugate and cut locus points with the ensuing formations of past light cone caustics, see Fig.\ref{<fig3>}. The presence of multiple images of the same astrophysical source on the observer's celestial sphere, gravitational lensing\cite{Perlick}, and large image distortions offer a dramatic demonstration of this behavior 
and also show that data gathered on $\mathbb{CS}_z$ provide explicit geometric information. In particular, they characterize an observable two-dimensional metric $h_{(z)}$ on the celestial sphere, a metric that coexists with the 2-sphere round metric $\widetilde{h}(\mathbb{S}^2)$ naturally defined on $\mathbb{CS}_z$,  and that, as the redshift varies, allows us to track down the actual spacetime geometry along the past light cone and of the null geodesics distortion effect due to inhomogeneities (the symbols $h_{(z)}$and $\widehat{h}_{(\widehat{z})}$, denoting the metrics on celestial spheres, should not be confused with the adimensional Hubble constant parameter $\mathrm{h}$). 
As stressed above, magnification,  distortions, and multiple imaging are the rule rather than the exception, and the information null geodesics convey to the physical celestial sphere $\mathbb{CS}_z$ is quite corrupted by the presence of the gravitational inhomogeneities that they encounter evolving from the astrophysical sources to the observer. Since we are not interested in the details and the nature of the specific strong lensing events, one can safely make the working assumption that $(\mathbb{CS}_z,\,h_{(z)})$ is topologically a 2-sphere with a potentially fractal distance (Lipschitz) geometry. 
Astrophysical data are similarly gathered by the reference FLRW observer on $\widehat{\mathbb{CS}}_{\widehat{z}}$ where they induce a metric $\widehat{h}_{(\widehat{z})}$ that gives the FLRW celestial sphere $(\widehat{\mathbb{CS}}_{\widehat{z}},\,\widehat{h}_{(\widehat{z})})$ the ($\widehat{z}$- scaled) round geometry of a sphere. Whereas the celestial sphere  $(\mathbb{CS}_z,\,h_z)$  has a rather rough landscape, data on $(\widehat{\mathbb{CS}}_{\widehat{z}},\,\widehat{h}_{(\widehat{z})})$ are visualized and interpreted in terms of the smooth FLRW spacetime geometry. At the common observation event $p$ and assuming, for simplicity, that the relative velocity $v(p)$ of the two observers vanishes, the area elements $d\mu_{{h}_{({z})}}$ and    $d\mu_{\widehat{h}_{(\widehat{z})}}$ of  $(\mathbb{CS}_z,\,h_z)$  and $(\widehat{\mathbb{CS}}_{\widehat{z}},\,\widehat{h}_{(\widehat{z})})$ provide the cross-sectional areas of the given source as seen by the physical and the FLRW observers in their local rest space. In particular, if the source subtends the observer's visual solid angle $d\mu_{\mathbb{S}^2}$ on the unit sphere, and if the observers know the intrinsic size of the given source, then the  physical area distance ${D}(\zeta_{(\widehat{z})}(y))$ and the FLRW area distance $\widehat{D}(y)$, between the observation event $p$ and the  astrophysical sources located at the celestial coordinates $y$ on $\widehat{\mathbb{CS}}_{\widehat{z}}$ and $\zeta_{(\widehat{z})}(y)$ on $\mathbb{CS}_z$, are defined by
$d\mu_{{h}_{({z})}}\,=\,{D}^2(\zeta_{(\widehat{z})}(y))\,d\mu_{\mathbb{S}^2}$ and 
$d\mu_{\widehat{h}_{(\widehat{z})}}\,=\,\widehat{D}^2(y)\,d\mu_{\mathbb{S}^2}$, see.
Area distances are, in principle, measurable quantities whenever the source is an astrophysical object of known type. In this connection, it is worthwhile to stress that the celestial spheres area elements $d\mu_{{h}_{({z})}}$ and  $d\mu_{\widehat{h}_{(\widehat{z})}}$ are related to the intrinsic size of the source through the mapping action of the null geodesic flow reaching the observer from the source, (technically, there is a non-trivial action of the Lipschitz exponential map associated with the null geodesic flow). ${D}(\zeta_{(\widehat{z})}(y))$ and  $\widehat{D}(y)$ are distinct from each other. In particular, ${D}(\zeta_{(\widehat{z})}(y))$ has a memory of all the distortions that affect null geodesic propagation in the pre-homogeneity region. For this reason, the geometric landscape offered by the celestial sphere $({\mathbb{CS}}_{{z}},\,{h}_{({z})})$ is quite different from the one described by  
$(\widehat{\mathbb{CS}}_{\widehat{z}},\,\widehat{h}_{(\widehat{z})})$, and their comparison requires a rather delicate approach\cite{CarFam}.
 As in cartography, where a smooth geographical globe is compared with a physical 3-dimensional scaled rendering of the actual Earth surface, one can exploit\cite{CarFam} the fact that the underlying geometries are conformally 
related,  
$\zeta_{(_{\widehat{z}})}^*\,{{h}_{({z})}}\,=\,\Phi^2(\zeta_{(_{\widehat{z}})})\,{\widehat{h}_{(\widehat{z})}}$, where $\zeta_{(_{\widehat{z}})}^*$ denotes the action (the pull-back) of the Lorentz aberration map connecting $\widehat{\mathbb{CS}}_{\widehat{z}}$ with ${\mathbb{CS}}_{{z}}$. The Poincar\'e-Koebe uniformization theorem for 2-dimensional surfaces is the basic tool that allows us to conformally represent bumpy 2-spheres, with a complex pattern of raised relief and large landforms (like the Earth's surface) over a smooth and round geographical globe. In the same spirit, the conformal factor  $\Phi(\zeta_{(_{\widehat{z}})})$ does the job here and keeps track of what happens in the pre-homogeneity region (Fig.\ref{<fig5>}). Not surprisingly, it can be related to the area distances according to 
$ \Phi^2(\zeta_{(_{\widehat{z}})})\,=\,{D}^2(\zeta_{(_{{z}})})/\widehat{D}^2$, a connection that characterizes\cite{CarFam, CarFam2, Carfora3} the  functional which, at the given redshift $\widehat{z}$, allows us to compare the FLRW celestial sphere $\widehat{\mathbb{CS}}_{\widehat{z}}$ and the physical celestial sphere $\mathbb{CS}_{z}$,\,\emph{i.e.}
\begin{equation}
\label{Efunctn}
E_{\widehat{\mathbb{C\,S}}_{\widehat{z}},\;{\mathbb{C\,S}}_{z}}[\zeta_{(\widehat{z})}]\,:=\,\int_{\widehat{\mathbb{CS}}_{\widehat{z}}}\left({\Phi}(\zeta_{(_{\widehat{z}})})\,-\,1 \right)^2\,d\mu_{\hat{h}_{(\widehat{z})}}\,=\,
\int_{\widehat{\mathbb{CS}}_{\widehat{z}}}\,
\left[\frac{D_{{z}}(\zeta_{(\widehat{z})})\,-\,\widehat{D}_{\widehat{z}}}{\widehat{D}_{\widehat{z}}}\right]^2
\,d\mu_{\hat{h}_{(\widehat{z})}}\,.
\end{equation} 
This functional belongs to an important class of scale-dependent functionals that are efficiently used in image visualization\cite{HassKoehl}. 
From a geometric analysis point of view, it can be minimized\cite{CarFam} with respect to the choice of the aberration map $\zeta_{(_{\widehat{z}})}$, providing a scale-dependent  "distance functional" $E_{\widehat{\mathbb{C\,S}}_{\widehat{z}},\;{\mathbb{C\,S}}_{z}}\,:=\inf_{\zeta_{(\widehat{z})}}\,E_{\widehat{\mathbb{C\,S}}_{\widehat{z}},\;{\mathbb{C\,S}}_{z}}[\zeta_{(\widehat{z})}]$\;
between the FLRW celestial sphere $(\widehat{\mathbb{CS}}_{\widehat{z}},\,\widehat{h}_{(\widehat{z})})$  and the physical celestial sphere $({\mathbb{CS}}_{{z}},\,{h}_{({z})})$.  In particular, it vanishes if and only if the two celestial spheres are isometric.

\section{Memories of late-epoch inhomogeneities}
In terms of the sky averages of the area distances relative fluctuations,  
$\delta^{(n)}_{\widehat{\mathbb{C\,S}}_{\widehat{z}},\;{\mathbb{C\,S}}_{\widehat{z}}}:=\frac{1}{4\pi}
\int_{\mathbb{S}^2}
[\frac{D_{z}-\widehat{D}_{\widehat{z}}}{\widehat{D}_{\widehat{z}}}]^n
d\mu_{\mathbb{S}^2}$,\; with $n=1,2$, the functional $E_{\widehat{\mathbb{C\,S}}_{\widehat{z}},\;{\mathbb{C\,S}}_{z}}$ can be equivalently rewritten in adimensional form as      
\begin{equation}
\frac{E_{\widehat{\mathbb{C\,S}}_{\widehat{z}},\;{\mathbb{C\,S}}_{z}}}{4\pi\,\widehat{D}^{2}_{(\widehat{z})}}\,=\,
\delta^{(2)}_{[\widehat{\mathbb{C\,S}}_{\widehat{z}},\;{\mathbb{C\,S}}_{\widehat{z}}]}\,
\,=\,
\frac{A({\mathbb{C\,S}}_{z})\,-\,A(\widehat{\mathbb{C\,S}}_{\widehat{z}})}{4\pi\,\widehat{D}^{2}_{(\widehat{z})}}\,-2\,\delta^{(1)}_{[\widehat{\mathbb{C\,S}}_{\widehat{z}},\;{\mathbb{C\,S}}_{\widehat{z}}]}\,,
\end{equation} 
where $A(\widehat{\mathbb{C\,S}}_{\widehat{z}})$ and $A({\mathbb{C\,S}}_{z})$ are the areas of the celestial spheres  $(\widehat{\mathbb{CS}}_{\widehat{z}},\,\widehat{h}_{(\widehat{z})})$ and $({\mathbb{CS}}_{{z}},\,{h}_{({z})})$.
According to the Lorentzian version of the Bertrand-Puiseaux formulas\cite{Berthiere} (familiar in Riemannian geometry when defining curvatures in terms of the surface areas and volumes of small geodesic balls), these areas are related to the spacetime scalar curvature in the pre-homogeneity region, and for  $\widehat{z}\rightarrow 0$ we have the limiting behavior\cite{CarFam}
\begin{equation}
\label{limitMem}
\lim_{\widehat{z}\,\longrightarrow\,0}\,
\frac{24\,(1+q_0)^2\,\left(
\delta^{\;(2)}_{\widehat{\mathbb{C\,S}}_{\widehat{z}},\;{\mathbb{C\,S}}_{{z}}}+2\delta^{\;(1)}_{\widehat{\mathbb{C\,S}}_{\widehat{z}},\;{\mathbb{C\,S}}_{{z}}}\right)}{\left(1\,+\,\widehat{z}\right)^{\;2}\,\ln^2\left[1\,+\,(1+q_0)\widehat{z}\right]}\,
=\,\frac{\widehat{\mathrm{R}}(p)\,-\,\mathrm{R}(p)}{12\,H_0^2}\,,
\end{equation}  
where $q_0\,\simeq\,-\,0.55$ is the present value of the deceleration parameter, and
where $\mathrm{R}(p)$ and $\widehat{\mathrm{R}}(p)$ respectively are the scalar curvatures of the physical and of the reference FLRW spacetimes (evaluated at the common observation point $p$). This relation holds asymptotically (up to $\mathcal{O}(\widehat{z}^2)$ terms) also in a neighborhood of $\widehat{z}=0$, say in the low-moderate redshift range. Thus, even if it would have been safe to use, in (\ref{limitMem}), the standard Hubble formula  
$(c)\widehat{z}=H_0\,r$ with no role for the deceleration parameter $q_0$, the asymptotics associated with (\ref{limitMem}) that we use below probes the range $0<\widehat{z}\lesssim 10^{-2}$, where the actual Hubble constant is better approximated by $H_0[1+(1+q_0)z]$.   
It is important to keep track of the different roles that the relative fluctuation terms $\delta^{\;(1)}$ and 
$\delta^{\;(2)}$ play in the relation (\ref{limitMem}).\;In line with the FLRW point of view, let us assume that, as the redshift $\widehat{z}$ varies, the relative fluctuation of the area distance $(D_{z}-\widehat{D}_{\widehat{z}})/\widehat{D}_{\widehat{z}}$ is a normally distributed random variable on the celestial spheres $\widehat{\mathbb{C\,S}}_{\widehat{z}}$, and the term $\delta^{\;(1)}$ is the associated sample mean over the solid angle. The angular sample variance of $(D_{z}-\widehat{D}_{\widehat{z}})/\widehat{D}_{\widehat{z}}$ is given by
\begin{equation}
\frac{1}{4\pi}
\int_{\mathbb{S}^2}
\left[\frac{D_{z}-\widehat{D}_{\widehat{z}}}{\widehat{D}_{\widehat{z}}}\,-\,\delta^{\;(1)}\right]^2
d\mu_{\mathbb{S}^2}\,=\,\delta^{\;(2)}\,-\,\left(\delta^{\;(1)} \right)^2\,,
\end{equation}
and the independence of the sample mean and sample variance for normally distributed data implies that also $\delta^{\;(1)}$ and  $\delta^{\;(2)}$ are independent. To see the different roles that $\delta^{\;(1)}$ and $\delta^{\;(2)}$ play in (\ref{limitMem}),  let $\widehat{\rho}_{\widehat{z}}$ denote the matter density in the FLRW reference spacetime at the given $\widehat{z}$. Typically, at late epoch, it is customary to assume a pure matter scheme and interpret $\widehat{\rho}_{\widehat{z}}$ as an average of sorts of the matter density distribution over large scales. Similarly, to avoid dealing with the detailed vagaries of the matter distribution, the physical observer considers the sky average ${\rho}_{{z}}$ of the actual matter density in the pre-homogeneity region at redshift $z$. As $z$ varies, ${\rho}_{{z}}$  statistically fluctuates  with respect to $\widehat{\rho}_{\widehat{z}}$, and the small $\widehat{z}$ asymptotics associated with (\ref{limitMem}), coupled with Einstein's equations, naturally puts forward three distinct regimes. To discuss them, let us recall that the  (trace) Einstein equations associated with  the physical and the reference FLRW spacetime provide (at the related $\widehat{z}$ and $z$),    $\widehat{\mathrm{R}}_{\widehat{z}}\,-\,\mathrm{R}_z=4(\widehat{\Lambda}^{(FLRW)}-\Lambda^{(phys)})+8\pi(\widehat{\rho}_{\widehat{z}}-\rho_z)$, where  $\widehat{\Lambda}^{(FLRW)}$ and $\Lambda^{(phys)}$ denote the FLRW cosmological constant and the actual cosmological constant of the physical spacetime, that we assume not necessarily coinciding with each other. The statistical independence of $\delta^{\;(1)}_{\widehat{\mathbb{C\,S}}_{\widehat{z}},\;{\mathbb{C\,S}}_{{z}}}$  and $\delta^{\;(2)}_{\widehat{\mathbb{C\,S}}_{\widehat{z}},\;{\mathbb{C\,S}}_{{z}}}$, \; and the structure of the   trace Einstein equation comprising the fluctuating density contrast term and the cosmological constants term allows us to factorize the redshift asymptotics associated with (\ref{limitMem}) according to
\begin{equation}
\label{mattertermsCorr}
\frac{\left(\widehat{\rho}(p)-\rho(p)\right)}{\widehat\rho(p)}\,
\Omega^{(FLRW)}_m\,=\,
\frac{192\,\delta^{\;(1)}_{\widehat{\mathbb{C\,S}}_{\widehat{z}},\;{\mathbb{C\,S}}_{{z}}}\,(1+q_0)^2}{\left(1\,+\,\widehat{z}\right)^{\;2}\,\ln^2\left[1\,+\,(1+q_0)\widehat{z}\right]}
\,+\,\ldots\,,
\end{equation}
\begin{equation}
\label{OmegatermsCorr}
\Omega^{(FLRW)}_\Lambda\,=\,\Omega^{(phys)}_\Lambda\,+\,
\frac{24\,\delta^{\;(2)}_{\widehat{\mathbb{C\,S}}_{\widehat{z}},\;{\mathbb{C\,S}}_{{z}}}\,(1+q_0)^2}{\left(1\,+\,\widehat{z}\right)^{\;2}\,\ln^2\left[1\,+\,(1+q_0)\widehat{z}\right]}\,,
\end{equation}
\\
\noindent
where in the fluctuating relation (\ref{mattertermsCorr}) $...$ stands for $\mathcal{O}(\widehat{z}^2)$ corrections, and where we have introduced the standard normalization to $H_0^2$ (we use geometrical units $c=1=G/c^2$), by defining  $\Omega^{(FLRW)}_\Lambda:=\widehat{\Lambda}/{3\,H_0^2}$,\, $\Omega^{(phys)}_\Lambda:=\Lambda/{3\,H_0^2}$,\, and \,$\Omega^{(FLRW)}_m:=8\pi\widehat{\rho}/{3\,H_0^2}$. These relations imply the following scenario as seen from the physical and the FLRW observers located at the barycenter of a galaxy cluster.

\section{Does pre-homogeneity contribute to the Cosmological constant?}
Galaxy clusters have an average matter density\,$\rho_z$, which is quite larger than the FLRW average density $\widehat{\rho}_{\widehat{z}}$. Gravitational instabilities dominate, and the neighborhood of a physical observer located at a cluster barycenter $p$ is decoupled from cosmological expansion. We are dealing with a virialized structure where redshifts are due to the local relative motion and local gravitational fields, and the physical observer refers to the FLRW redshift $\widehat{z}$ only to characterize the peculiar velocities of her neighboring galaxies with respect to the idealized FLRW Hubble flow. The situation changes as we gather our astrophysical data from the outskirts of the cluster, where the infalling galaxies couple to the cosmological expansion. The radius of this region defines the \emph{physical radius} of the cluster. A typical cluster has a physical radius of $\sim\,3\,Mpc$; thus, according to the FLRW observer, a redshift of the order $\widehat{z}\sim 10^{-4}$ marks the \emph{inner cosmological boundary} of the pre-homogeneity region where gravitational instabilities interact significantly with the FLRW background cosmological expansion. Inhomogeneities, in the form of the inner core density contrast $(\widehat{\rho}(p)-\rho(p))/\widehat\rho(p)$,  leave their imprint on this boundary through two related mechanisms: \emph{(i)}\; A direct geometrical alteration of signal propagation that via the relation (\ref{mattertermsCorr}) 
is described by the (sky averaged) relative fluctuation of the area distance $\delta^{\;(1)}_{\widehat{\mathbb{C\,S}}_{\widehat{z}},\;
{\mathbb{C\,S}}_{{z}}}$; \; and \emph{(ii)} A positive contribution to the FLRW cosmological constant $\widehat{\Lambda}^{(FLRW)}$ due, according to the relation (\ref{OmegatermsCorr}), to the mean square relative fluctuation of the area distance $\delta^{\;(2)}_{\widehat{\mathbb{C\,S}}_{\widehat{z}},\;
{\mathbb{C\,S}}_{{z}}}$, or, more expressively, to the comparison functional $E_{\widehat{\mathbb{C\,S}}_{\widehat{z}},\;
{\mathbb{C\,S}}_{{z}}}$. From the Cauchy-Schwarz inequality we get
$\delta^{\;(2)}_{\widehat{\mathbb{C\,S}}_{\widehat{z}},\;
{\mathbb{C\,S}}_{{z}}}\,\geq\,(\delta^{\;(1)}_{\widehat{\mathbb{C\,S}}_{\widehat{z}},\;
{\mathbb{C\,S}}_{{z}}})^2$, (where equality holds if and only if $\delta^{\;(2)}_{\widehat{\mathbb{C\,S}}_{\widehat{z}},\;
{\mathbb{C\,S}}_{{z}}}=0$), which implies that $E_{\widehat{\mathbb{C\,S}}_{\widehat{z}},\;
{\mathbb{C\,S}}_{{z}}}$  is strictly connected to the density contrast and is there to stay as long as there are density fluctuations around. This contribution to the cosmological constant may be potentially significant, and estimating its order of magnitude is worthwhile.\\
We start with the relative fluctuation 
term $\delta^{\;(1)}_{\widehat{\mathbb{C\,S}}_{\widehat{z}},\;
{\mathbb{C\,S}}_{{z}}}$. If we assume the typical cluster density contrast  $|\widehat{\rho}(p)-\rho(p)|/\widehat\rho(p)\sim\,10^6$, and take the \emph{Planck} value $\Omega^{(FLRW)}_m\sim 0.3$ for the average FLRW density\cite{Planck}, then as we approach the \emph{inner cosmological boundary} at $\widehat{z}\sim\ 10^{-4}$,  the relation (\ref{mattertermsCorr}) provides 
$|\delta^{\;(1)}_{\widehat{\mathbb{C\,S}}_{\widehat{z}},\;
{\mathbb{C\,S}}_{{z}}}|\sim\,10^{-5}$.\, A result that is in very good agreement with the estimates of $\delta^{\;(1)}$ obtained in this regime using FLRW perturbation theory\cite{Durrer, Fanizza, Heinesen, Obinna4}. From the inequality  
$\delta^{\;(2)}_{\widehat{\mathbb{C\,S}}_{\widehat{z}},\;
{\mathbb{C\,S}}_{{z}}}\,\geq\,(\delta^{\;(1)}_{\widehat{\mathbb{C\,S}}_{\widehat{z}},\;
{\mathbb{C\,S}}_{{z}}})^2$ we have 
$\delta^{\;(2)}_{\widehat{\mathbb{C\,S}}_{\widehat{z}},\;
{\mathbb{C\,S}}_{{z}}}\,\geq\, 10^{-10}$, thus from (\ref{OmegatermsCorr}) we get
\begin{equation}
\label{OmegatermsCorr2}
\frac{\Omega^{(FLRW)}_\Lambda\,-\,\Omega^{(phys)}_\Lambda}{\Omega^{(FLRW)}_\Lambda}\,=\,\left.
\frac{24\,\delta^{\;(2)}_{\widehat{\mathbb{C\,S}}_{\widehat{z}},\;{\mathbb{C\,S}}_{{z}}}\,(1+q_0)^2}{\Omega^{(FLRW)}_\Lambda\left(1\,+\,\widehat{z}\right)^{\;2}\,\ln^2\left[1\,+\,(1+q_0)\widehat{z}\right]}\right|_{\widehat{z}\sim 10^{-4}}\,\geq\,\frac{1}{3}.
\end{equation}
Thus, as we approach the \emph{inner cosmological boundary} in the pre-homogeneity region, the correction $\Omega^{(corr)}_\Lambda\,:=\,\Omega^{(FLRW)}_\Lambda\,-\,\Omega^{(phys)}_\Lambda$, to the cosmological constant due to the density contrast is of the same order of magnitude as the assumed FLRW contribution $\Omega^{(FLRW)}_\Lambda\simeq\,69\times 10^{-2}$, \emph{i.e.} explicitly
\begin{equation}
\Lambda^{(corr)}\,:=\,\Lambda^{(FLRW)}-\Lambda^{(phys)}\,\sim\,10^{-\,52}\,m^{-2}\,.
\end{equation}
If we assume a Copernican point of view, according to which every late-epoch 
FLRW fundamental observer is at the center of a pre-homogeneity region, then 
one is strongly tempted to set $\Lambda^{(Phys)}=0$, and to identify 
$\Lambda^{(FLRW)}$ with an average of sorts of the inhomogeneities induced 
correction terms  $\Lambda^{(corr)}$ experienced by the fundamental observers 
in their pre-homogeneity regions. Since inhomogeneities and the associated 
density contrast builds up as we approach the late-epoch observer, 
$\Lambda^{(corr)}$ is a scale-dependent correction, as is manifest 
from (\ref{OmegatermsCorr}). This scale-dependence of $\Lambda^{(corr)}$ suggests a relation with the coincidence problem connecting the dominance of dark energy with the formation of structures.  In particular, as $\widehat{z}$ increases from 
the critical $\widehat{z}\sim\,10^{-4}$ to the other landmarks of the pre-homogeneity region, the supercluster scale ($\sim\,10\,Mpc$) at  $\widehat{z}\sim\ 10^{-3}$ and eventually the homogeneity scale 
$100\,h^{\,-\,1}\,Mpc$, \, at  $\widehat{z}\sim\ 10^{-2}$, we easily estimate that the corresponding correction term $\Omega^{(corr)}_\Lambda$ decreases as $\widehat{z}$ increases. To wit, in the small redshift range  $10^{-4}\lesssim\widehat{z}\lesssim 10^{-2}$, we get $\Omega^{(corr)}_\Lambda(\widehat{z}\sim\ 10^{-3})\sim 10^{-4}$,\, and  $\Omega^{(corr)}_\Lambda(\widehat{z}\sim\ 10^{-2})\sim 10^{-8}$. For larger redshifts, the asymptotics (\ref{OmegatermsCorr}) must be corrected with the higher order term, and it is quite delicate to manipulate for $\widehat{z}>1$.  

\section{Discussion}
In this paper, we have introduced a comparison functional in mainstream FLRW cosmology that allows the FLRW observer to compare her reference past light cone with the physical light cone characterized by the actual data she gathers. The functional originates from visualization techniques used in imaging, and it can be directly related to the relative fluctuations in area distance, a quantity that is observable and is intensively investigated in connection with the Hubble tension debate. At the late epoch, when the cosmological expansion decouples from the local virialized gravitational dynamics, our comparison functional keeps memory of the non-perturbative density contrast associated with gravitational instabilities. It provides a redshift-dependent positive contribution to the cosmological constant of the same order of magnitude as the FLRW cosmological constant itself. It is important to stress that regardless of its size, the presence of this correction is a rigorous mathematical consequence of the nature of the physical past light cone. It is present as soon as there is a density contrast. The result proven here can be interpreted, from the mainstream FLRW point of view, as the presence of an effective field associated with the presence of a pre-homogeneity region surrounding, in the late-epoch, each fundamental FLRW observer. This \emph{pre-homogeneity field} is scale-dependent (as a convenient scale, we used the FLRW redshift $\widehat{z}$), and it reaches its maximum in the pre-homogeneity region surrounding the given FLRW observer, where the complex network of structures couples with the background FLRW cosmological expansion. As we have shown, in these regions, typically identified with the physical radius of clusters, this effective field contributes as a gravitational constant of the same order of magnitude as the assumed $\Lambda^{(FLRW)}$.
Moreover, the fact that this effective observer-dependent field is scale-dependent may also provide an explanation of the coincidence problem,  according to which the expansion of the Universe appears to have started accelerating at the same epoch when complex nonlinear structures emerged. It is also important to stress that this approach goes a long way in realizing, in a relatively simple and effective way, the best-fit program\cite{EllisStoeger} advocated by G. F. R. Ellis and W. R. Stoeger. It  highlights a methodological scenario that is potentially significant for interpreting observations of the actual inhomogeneous Universe and that may lead to a new perspective on the dark energy problem, one of the fundamental unsolved issues in modern cosmology.

\bmhead{Acknowledgments}
We are grateful to T. Buchert and G.F.R. Ellis for valuable discussions that significantly improved a preliminary version of the paper.


\begin{thebibliography}{99}
\bibitem{Planck} Planck Collaboration, \textit{Planck 2018 results. 
VI. Cosmological parameters}, A\& A Vol. \textbf{641}, A6, (2020) (1-67)
DOI 10.1051/0004-6361/201833910.

\bibitem{SEH} W. R. Stoeger, G. F. R. Ellis and C. Hellaby, \textit{The relationship between continuum homogeneity and statistical homogeneity in cosmology}, Mon. Not. R. astr. Soc.  \textbf{226} (1987), 373-381.

\bibitem{Maartens} R. Maartens, \textit{Is the Universe homogeneous?} Philosophical Transactions of the Royal Society, \textbf{A 369} (2011), 5115-37.

\bibitem{Scrimgeour}  M. Scrimgeour M et al., \textit{The WiggleZ Dark Energy Survey: the transition to large-scale cosmic homogeneity} Mon. Not. R. Astr. Soc. 425 (2012) 116 [arXiv:1205.6812].

\bibitem{Dias} B. L. Dias, F. Avila, A. Bernui, \textit{Probing cosmic homogeneity in the Local Universe}, Monthly Notices of the Royal Astronomical Society \textbf{526}, (2023) 3219-3229.
 
\bibitem{Wiltshire} D.Wiltshire, \textit{Comment on "Hubble flow variations as a test for inhomogeneous cosmology"}, Astronomy and Astrophysics \textbf{624} (2019) A12.

\bibitem{EllisPadova} G. F. R. Ellis, \textit{Relativistic cosmology: its nature, aims and problems}, In General Relativity and Gravitation, ed. B Bertotti et al (Reidel, 1984), 215.

\bibitem{DarkBuchert} T. Buchert, \textit{Dark energy from structure: a status report}, GRG Journal \textbf{40}: 467 (2008) [arXiv:0707.2153]. 

\bibitem{BuchertSyksy} T. Buchert, S. R\"as\"anen, \textit{Backreaction in Late-Time Cosmology}, The Annual Review of Nuclear and Particle Science \textbf{62}, 57-79 (2012).

\bibitem{Ellis2} G.F.R. Ellis,  R. Maartens and M. A. H. MacCallum, \textit{Relativistic Cosmology}, Cambridge Univ. Press (2012).


\bibitem{Carfora0} T. Buchert, M. Carfora,  \textit{Regional averaging and scaling in relativistic cosmology}, Class. Quantum Grav. 19 6109 (2002)
 DOI 10.1088/0264-9381/19/23/314. 

\bibitem{Carfora1} T. Buchert, M. Carfora,  \textit{Cosmological Parameters Are Dressed},
Phys. Rev. Lett. 90, 031101, (2003). 
 
\bibitem{Buchert} T. Buchert, M. Carfora, G. F. R. Ellis,  E. W.  Kolb, M.  MacCallum, J. Ostrowski, S. R\"as\"anen, B. Roukema, L. Andersson,  A. Coley,  D. Wiltshire, \textit{Is there proof that backreaction of inhomogeneities is irrelevant in cosmology?}. Classical and Quantum Gravity \textbf{32} (21),  (2015).

\bibitem{Wald1} A. Ishibashi, and R. M. Wald, \textit{Can the Acceleration of Our Universe Be Explained by the Effects of Inhomogeneities?}, Class. Quant. Grav. \textbf{23} (2006) 235-250,
[arXiv:gr-qc/0509108].

\bibitem{Wald2} S. R. Green, and R. M. Wald, \textit{A new framework for analyzing the effects of small scale inhomogeneities in cosmology}, Phys. Rev. D \textbf{83} (2011) 084020,
[arXiv:1011.4920].

\bibitem{Wald3} S. R. Green, and R. M. Wald,\textit{Examples of backreaction of small scale inhomogeneities in cosmology} Phys. Rev. D \textbf{87} (2013) 124037,
[arXiv:1304.2318].
 
\bibitem{Durrer} C. Bonvin, C. Clarkson, D. Durrer, R. Maartens,  O. Umeh, \textit{ 
Do we care about the distance to the CMB? Clarifying the impact of second-order lensing}, Journal of Cosmology and Astroparticle Physics06(2015)050 DOI 10.1088/1475-7516/2015/06/050 [arXiv:1503.07831].

\bibitem{Fanizza} G. Fanizza, M. Gasperini, G. Marozzi, and G. Veneziano, \textit{A new approach to the propagation of
light-like signals in perturbed cosmological backgrounds}, JCAP 1508 (2015), no. 08 020,
[arXiv:1506.02003].

\bibitem{Heinesen} A. Heinesen, \textit{Multipole decomposition of the general luminosity distance 'Hubble law'- a new framework for observational cosmology}, Journal of Cosmology and Astroparticle Physics 05(2021)008 	
https://doi.org/10.1088/1475-7516/2021/05/008,  arXiv:2010.06534v2 [astro-ph.CO]. 

\bibitem{Obinna4} O. Umeh, C. Clarkson and R. Maartens, \textit{Nonlinear relativistic corrections to cosmological distances, redshift and gravitational lensing magnification. II  derivation}, Class. Quant.Grav. \textbf{31} (2014) 205001 [arXiv:1402.1933].

\bibitem{CarFam} M. Carfora, F. Familiari, \textit{A comparison theorem for cosmological lightcones}, Letters in Mathematical Physics, 111:53 (2021), https://doi.org/10.1007/s11005-021-01393-2

\bibitem{EllisStoeger} G. F. R. Ellis and W. R. Stoeger, \textit{The "Fitting Problem" in cosmology}, Class. Quant. Grav. \textbf{4}, 1697 (1987)

\bibitem{Gott} J. R. Gott, III, M. Juric, D. Schlegel, F. Hoyle, M. Vogeley, M. Tegmark, N. Bahcall, J. Brinkmann, \textit{A Map of the Universe},  Astrophys.J.624:463,(2005) (arXiv:astro-ph/0310571).

\bibitem{Hogg} D. W. Hogg, D. J. Eisenstein, M. R. Blanton, N. A. Bahcall, J. Brinkmann, J. E. Gunn, and D. P. Schneider, \textit{Cosmic Homogeneity Demonstrated with Luminous Red Galaxies}, The Astrophysical Journal, Vol. 624, (2005) 54-58. 

\bibitem{KolbMarraMatarrese} E. W. Kolb, V. Marra, S. Matarrese, \textit{Cosmological background solutions and cosmological backreactions}, Gen.Rel.Grav. \textbf{42} (2010) 1399-1412.

\bibitem{Perlick} V. Perlick, \textit{Gravitational Lensing from a Spacetime Perspective}, Living Reviews in Relativity, Livingreviews.org/lrr-2004-9.

\bibitem{Carfora3} M. Carfora, A. Marzuoli, \textit{Einstein Constraints and Ricci Flow:
A Geometrical Averaging of Initial Data Sets}, Mathematical Physics Studies (MPST), Springer (2023) ISBN: 978-981-19-8539-3.
 
\bibitem{CarFam2} M. Carfora, F. Familiari, \textit{A Scale-ependent Distance Functional between Past Light Cones in Cosmology}, Universe 2023, 9(1), 25; https://doi.org/10.3390/universe9010025 - 30 Dec 2022.

\bibitem{HassKoehl} J. Hass and P. Koehl, \textit{Comparing shapes of genus-zero surfaces},  Journal of Applied and Computational Topology, Vol. 1, (2017) 57-87.

\bibitem{Berthiere} C. Berthiere, G. Gibbons, and S. N. Solodukhin, \textit{Comparison theorems for causal diamonds}, Phys. Rev. D \textbf{92}, 064036 (2015). 
\end{thebibliography}
\end{document}